\documentclass{an}
\usepackage{graphicx}
\usepackage{times}
\Volume{000}             
\Year{0000}              
\Month{00}               
\Pagespan{000}{000}      

\begin{document}
\lhead[\thepage]{V. Joergens: Rotational Periods of very Young Brown Dwarfs in Cha\,I}
\rhead[Astron. Nachr./AN~{\bf XXX} (200X) X]{\thepage}
\headnote{Astron. Nachr./AN {\bf 32X} (200X) X, XXX--XXX}

\title{Rotational Periods of very Young Brown Dwarfs in Cha\,I}

\author{V. Joergens\inst{1} \and
Matilde Fernand\'ez\inst{2} \and Ralph Neuh\"auser\inst{1} \and
Eike Guenther\inst{3}}
\institute{
Max-Planck-Institut f\"ur Extraterrestrische Physik,
              Giessenbachstrasse 1, D-85748 Garching, Germany
\and
Instituto de Astrof\'{\i}sica de Andaluc\'{\i}a
(CSIC),
              Apdo. 3004,
              E-18080 Granada,
              Spain   
\and
Th\"uringer Landessternwarte, Tautenburg, 
Karl-Schwarzschild-Observatorium,
              Sternwarte 5,
              D-07778 Tautenburg,
              Germany
}
\date{Received {\it date will be inserted by the editor}; 
accepted {\it date will be inserted by the editor}} 

\abstract{A photometric monitoring campaign of brown dwarfs in the Cha\,I 
star forming region in the i and R band 
revealed significant periodic variations of the three M6.5--M7 type brown dwarf candidates 
Cha\,H$\alpha$\,2, 3 and 6 (Joergens et al. 2002).
These are the first rotational periods for very young (1--5\,Myr) 
brown dwarfs and among the first rotational periods for brown dwarfs at all.
The relatively long periods of 2.2 to 3.4 days as well as 
$v \sin i$ values (Joergens \& Guenther 2001) indicate that 
our targets are moderately fast
rotators in contrast to very rapidly rotating old brown dwarfs.
The periods for the Cha\,I brown dwarf candidates 
provide valuable data points in an as yet, in terms
of rotational characteristic, almost unexplored region of the age-mass 
diagram.
A comparison with rotational properties
of older brown dwarfs indicates that most of the acceleration
during the contraction phase takes place within the first 30\,Myr or less of the lifetime 
of a brown dwarf.
We have also determined periods for the two M5--M5.5 type
very low-mass stars B\,34 and CHXR\,78C.
}
\correspondence{viki@mpe.mpg.de}

\maketitle

\section{Data acquisition and period search}
Photometric i and R band data have been obtained
with DFOSC at the Danish 1.5\,m telescope at ESO. 
We performed aperture photometry and calculated differential magnitudes 
by means of carefully chosen reference stars in the field.

A search for periodicity in the obtained light curves has been
performed with the string-length method (Dworetsky 1983).
This method is ideally suited for unevenly spaced data as in our case. 
The algorithm phase folds the data with a trial period and calculates
the string length between successive data points. This is done for all periods
within a given period range. The period which generates the minimum
string length is the most likely period.

\section{Tracing the rotation by means of spots}
Young stars are kown to exhibit often prominent cool spots on their 
surface, analog to sunspots but of much larger size.
They are thought to form as magnetic flux tubes rise to the atmosphere
and suppress convective energy transport.
Appearance and disappearance of spots
due to stellar rotation modulates the total brightness
of the star and leads to periodic light variations.
The measured photometrical period of a spotted star is 
therefore a direct tracer of its rotational period.

\section{Results}
\begin{figure}[h]
\resizebox{\hsize}{!}
{\includegraphics[width=\textwidth]{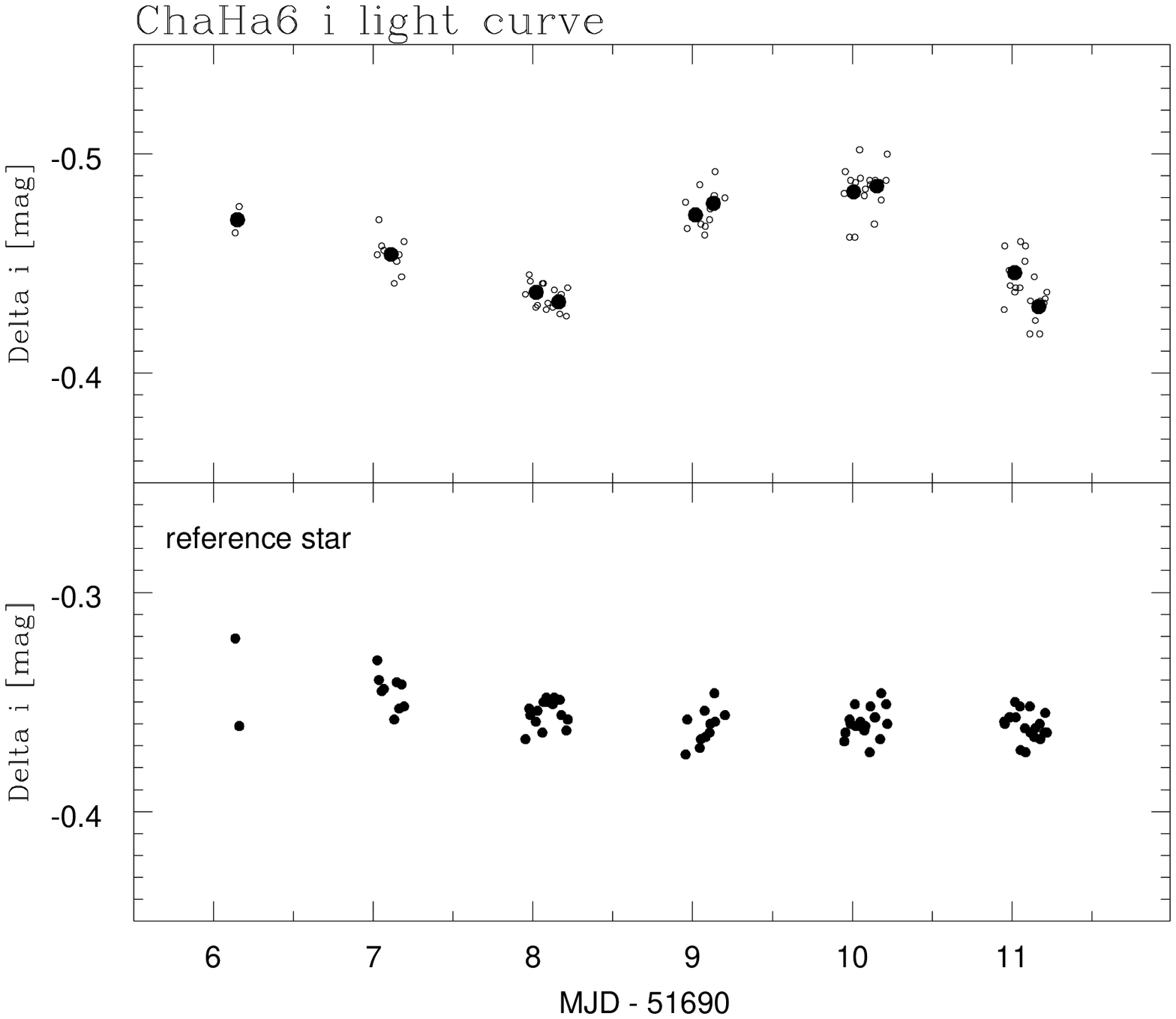}
\includegraphics[width=\textwidth]{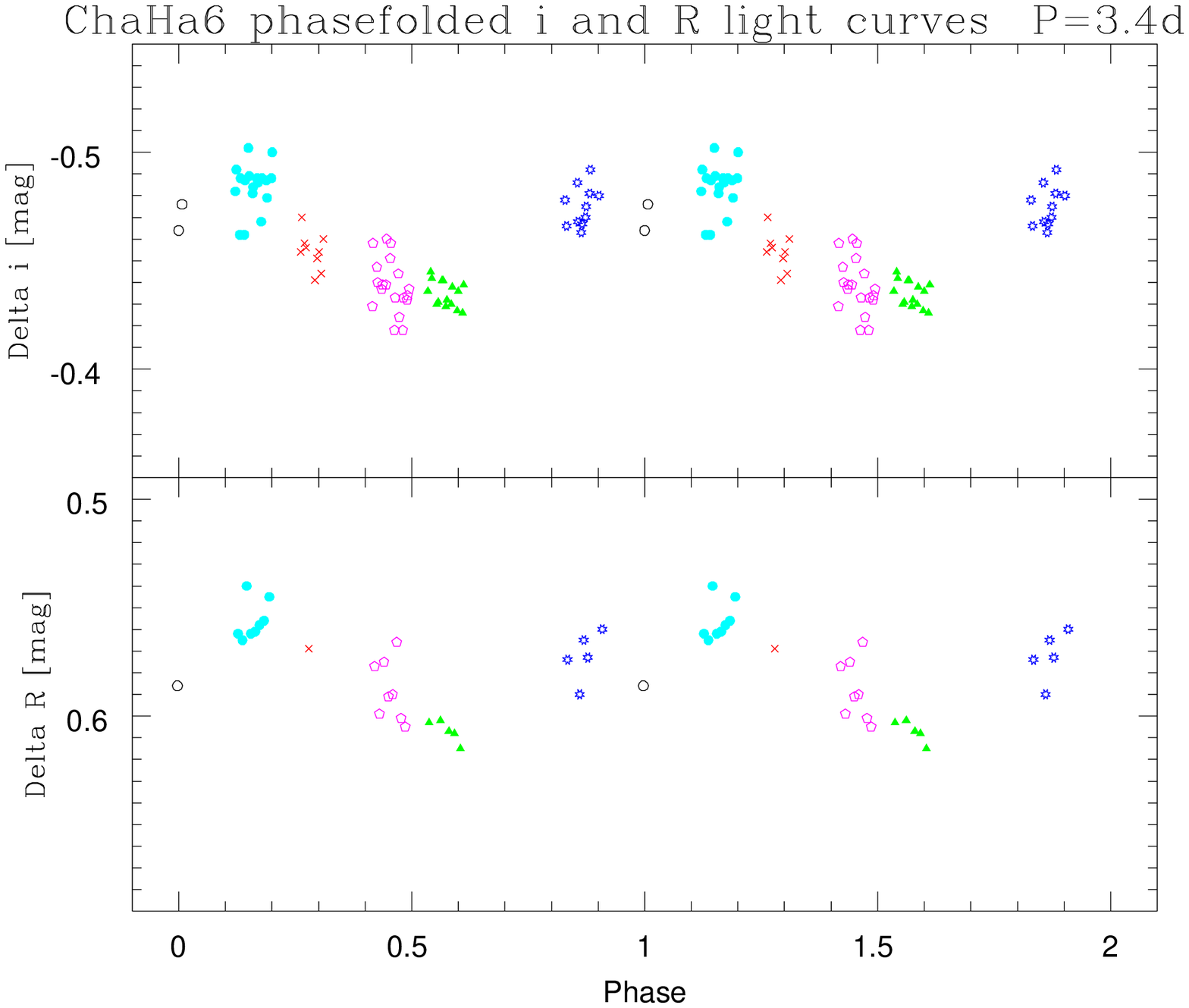}}
\caption{Phasefolded as well as unphased i-band light curve of 
Cha\,H$\alpha$\,6 
(M7, $\sim 0.05\,\mathrm{M}_{\odot}$, $\sim 0.68\,\mathrm{R}_{\odot}$)}
\label{cha6}
\end{figure}

\begin{figure}
\resizebox{\hsize}{!}
{
\includegraphics[width=\textwidth]{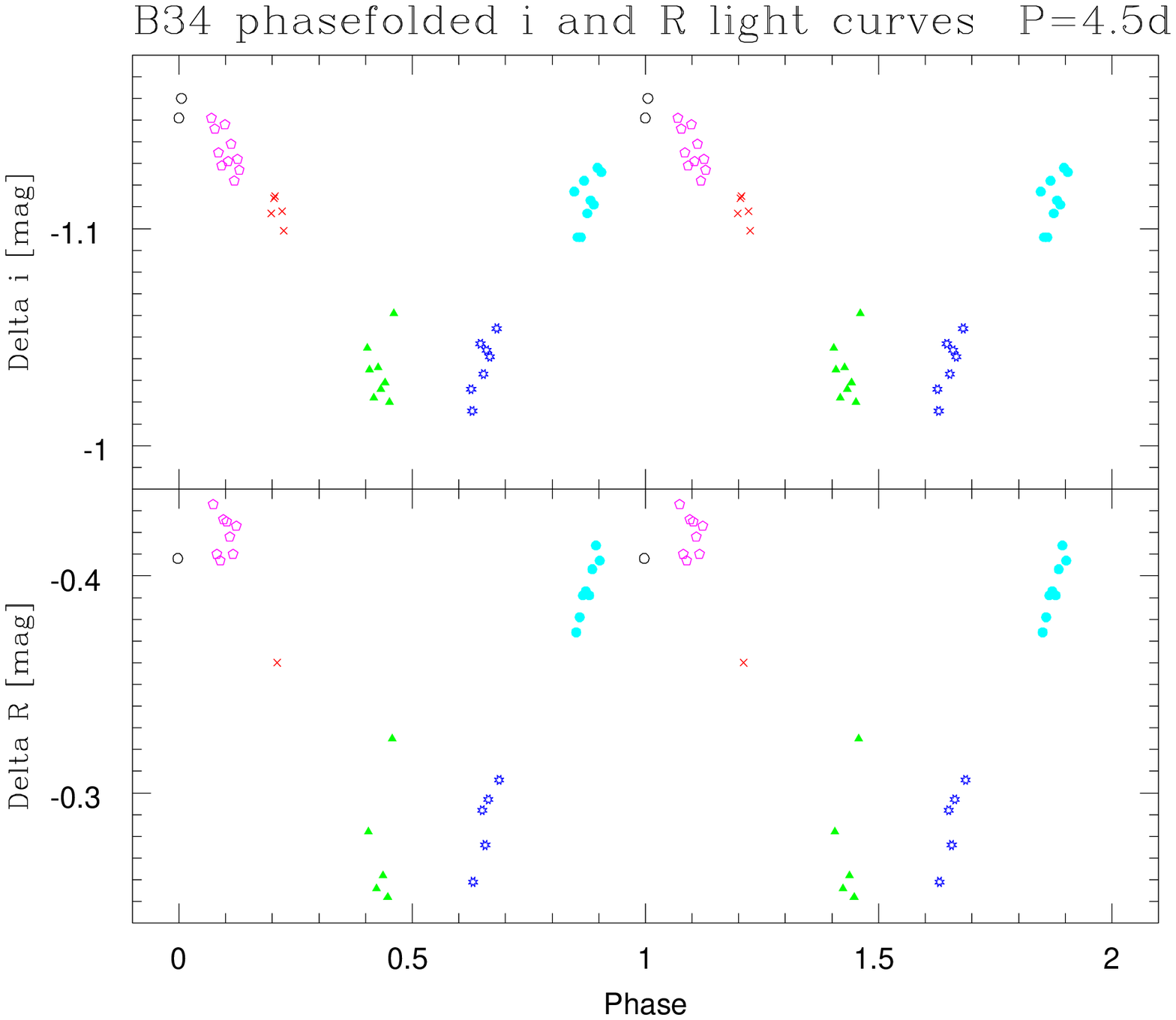}
\includegraphics[width=\textwidth]{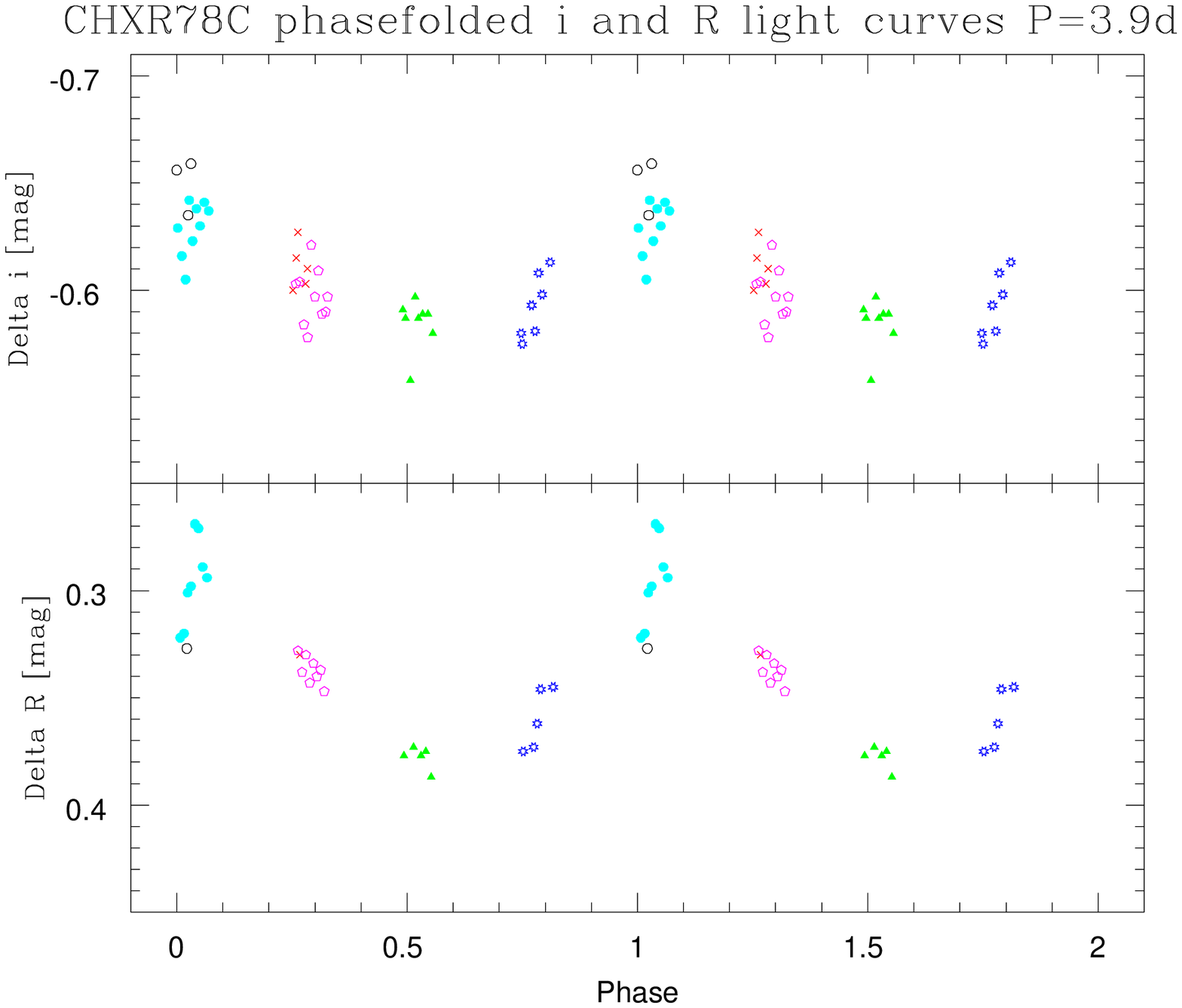}
}
\caption{Phasefolded light curves of the very low-mass T~Tauri stars
B34 (M5, $\sim 0.12\,\mathrm{M}_{\odot}$, $\sim 0.93\,\mathrm{R}_{\odot}$)
and CHXR\,78C 
(M5.5, $\sim 0.09\,\mathrm{M}_{\odot}$, $\sim 0.94\,\mathrm{R}_{\odot}$).}
\label{tts}
\end{figure}

\begin{figure}
\resizebox{\hsize}{!}
{
\includegraphics[width=\textwidth]{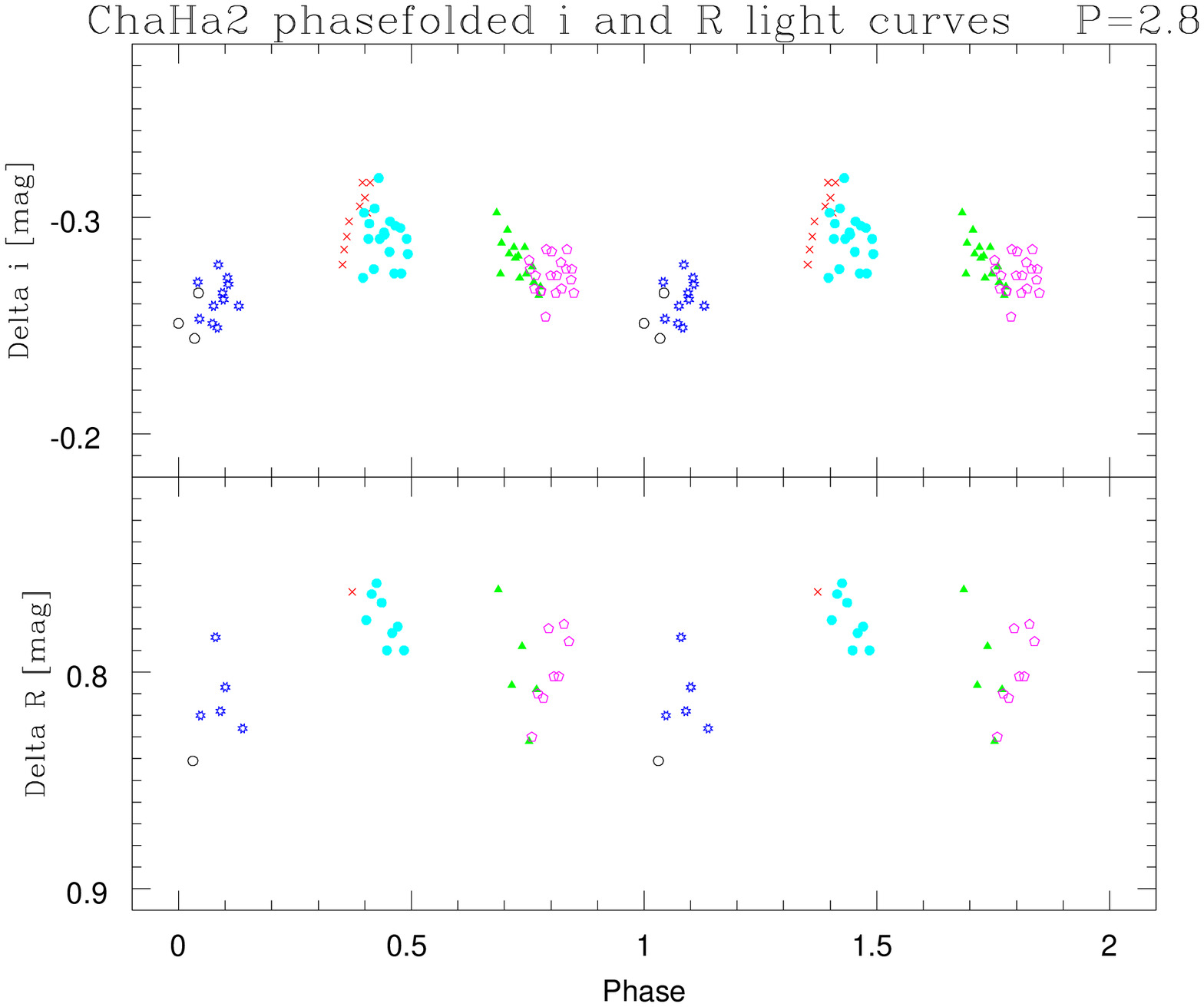}
\includegraphics[width=\textwidth]{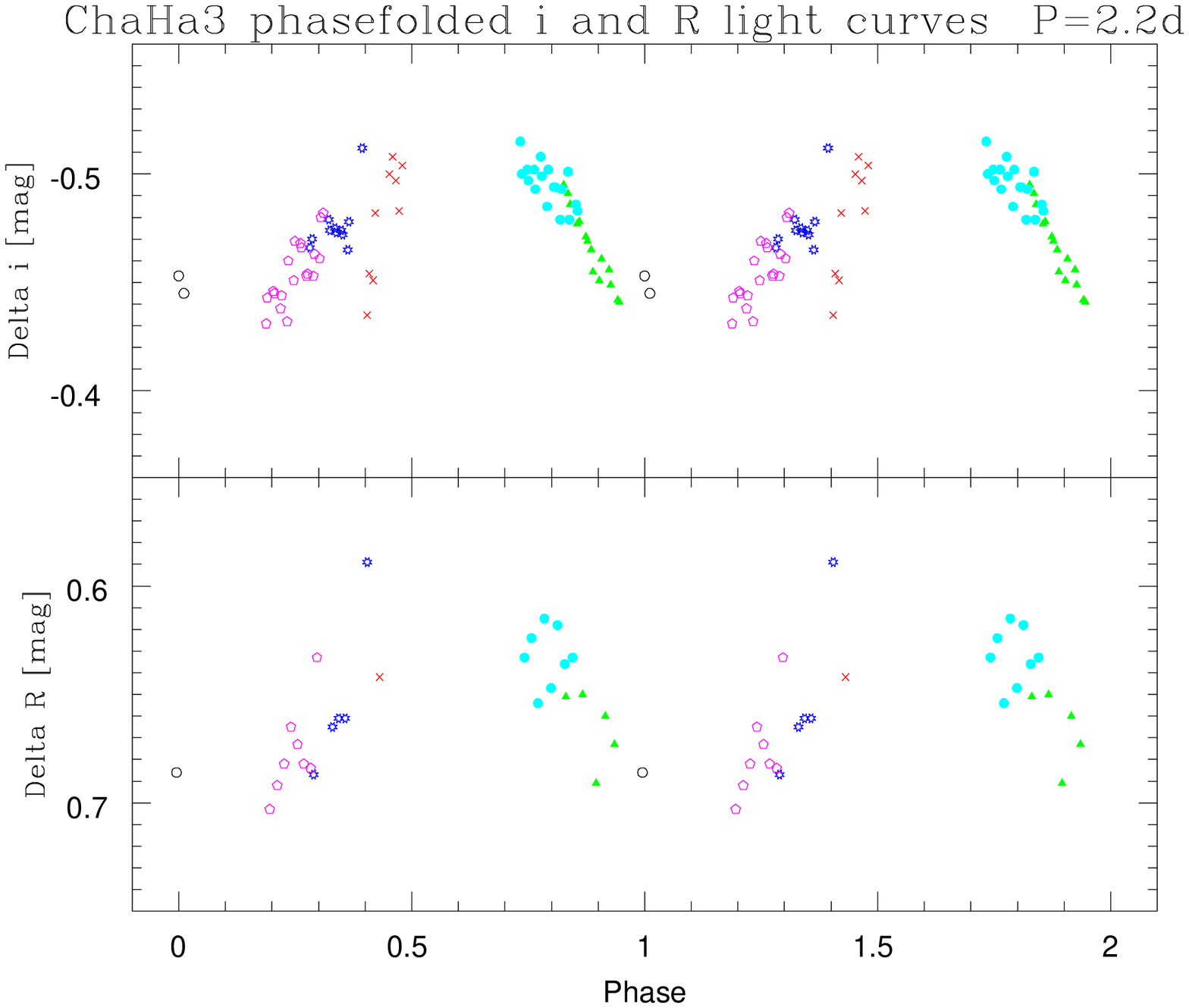}
}
\caption{Phasefolded light curves of the brown dwarf candidates
Cha\,H$\alpha$\,2 
(M6.5, $\sim 0.07\,\mathrm{M}_{\odot}$, $\sim 0.73\,\mathrm{R}_{\odot}$)
and Cha\,H$\alpha$\,3
(M7, $\sim 0.06\,\mathrm{M}_{\odot}$, $\sim 0.77\,\mathrm{R}_{\odot}$)
}
\label{cha2und3}
\end{figure}

We have measured significant periods for three brown dwarf candidates
(Cha\,H$\alpha$\,2, 3, 6) as well as for two very low-mass T~Tauri stars 
(B\,34, CHXR\,78C) within the range of 2.2\,d to 4.5\,d
and with i band amplitudes between 0.06\,mag and 0.14\,mag 
(Joergens et al. 2002).
See Table\,\ref{tab} for periods and $v \sin i$ measurements of the 
targets as well as Fig.\,\ref{cha6}, \ref{tts}, \ref{cha2und3} 
for the phasefolded light curves.

The observed periodic variations are interpreted as modulation of the 
	flux at the rotation period by magnetically driven surface 
	features on the basis of consistency with $v \sin i$ values as 
	well as (R-i) color variations typical for spots. 
	Furthermore, the temperatures even of the brown dwarfs in our sample
	are relatively high ($>$\,2800\,K) because they are very 
	young. Therefore,
	the atmospheric gas should be sufficiently ionized for the 
	formation of spots and the temperatures are
	too high for significant dust condensation and hence variability due
	to clouds.

\begin{table}[h]
\caption{
{\bf First rotation rates of very young brown dwarfs.}
Rotational periods $P_\mathrm{rot}$ (Joergens et al. 2002)
and measurements of the projected rotational velocity
$v \sin i$ based on UVES (VLT) spectra 
(Joergens \& Guenther 2001) for bona fide and candidate
brown dwarfs in Cha\,I with an age of 1--5\,Myr.
Spectral types SpT from Comer\'on et al. (2000).
}
\label{tab}
\begin{center}
\begin{tabular}{l|lcc}
\hline
object & SpT & $P_\mathrm{rot}$[d] & $v \sin i$[km/s] \\
\hline
Cha\,H$\alpha$\,1  & M7.5 & --  &  7.6$\pm$2.2 \\
Cha\,H$\alpha$\,2  & M6.5 & 2.8 & 12.8$\pm$1.2 \\
Cha\,H$\alpha$\,3  & M7   & 2.2 & 21.0$\pm$1.6 \\
Cha\,H$\alpha$\,4  & M6   & --  & 18.0$\pm$2.3 \\
Cha\,H$\alpha$\,5  & M6   & --  & 15.4$\pm$1.8 \\
Cha\,H$\alpha$\,6  & M7   & 3.4 & 13.0$\pm$2.8 \\
Cha\,H$\alpha$\,7  & M8   & --  & $\leq 10 $   \\
Cha\,H$\alpha$\,8  & M6.5 & --  & 15.5$\pm$2.6 \\
Cha\,H$\alpha$\,12 & M7   & --  & 25.7$\pm$2.6 \\
\hline
\end{tabular}
\end{center}
\end{table}

\section{Evolution of angular momentum \\in the substellar regime}
Old brown dwarfs are very rapid rotators 
having rotational velocities $v \sin i$ up to 60\,km\,s$^{-1}$
(e.g. Basri et al. 2000) and rotational periods shorter than one day 
(Bailer-Jones \& Mundt 2001, Mart\'\i n et al. 2001, 
Eisl\"offel \& Scholz 2001 and Clarke et al. 2002).
This can be explained by the fact that after the 
acceleration caused by contraction on the Hayashi track, 
there is no braking due to winds 
at the low masses of brown dwarfs, in contrast to stars.
However, our measure\-ments showed that extremely young brown dwarfs
have significantly longer rotational periods (2 to 3 days)
indicating that they are still in an early contracting stage and/or they
have suffered until very recently a braking due to their interaction
with an accretion disk. 
A comparison of our rotation periods at 1--5\,Myr 
and those in the literature at $>$\,36\,Myr
suggests that most of the acceleration takes place in
the first 30\,Myr or less of the lifetime of a brown dwarf.
(The mark at 36\,Myr ist set by rotational periods for brown dwarf candidates
in the cluster IC\,4665 by Eisl\"offel \& Scholz 2001.)


\end{document}